\documentclass[12pt,preprint]{aastex6}
\usepackage{exscale}
\usepackage{Picins}
\usepackage{amsmath}
\usepackage{color}
\usepackage{rotating}
\usepackage{graphicx}
\usepackage{epstopdf}
\usepackage{float}
\usepackage{amsfonts}
\usepackage{amssymb}
\usepackage{hyperref}
\usepackage{natbib}
\usepackage{caption2}
\usepackage{subfigure}
\usepackage{overpic}
\usepackage{ulem}
\usepackage{xcolor}

\renewcommand{\thefootnote}{\alph{footnote}}

\shorttitle{First determination of 2D speed distribution within the bodies of Coronal Mass Ejections}
\shortauthors{Ying et al.}

\begin{document}
\title{First determination of 2D speed distribution within the bodies \\
of Coronal Mass Ejections with cross-correlation analysis}
\author{Beili Ying\altaffilmark{1,2,}\footnote{This work was mainly carried out at INAF-Turin Astrophysical Observatory, via Osservatorio 20, 10025 Pino Torinese (TO), Italy.}, Alessandro Bemporad\altaffilmark{3}, Silvio Giordano\altaffilmark{3}, Paolo Pagano\altaffilmark{4}, Li Feng\altaffilmark{1,2}, Lei Lu\altaffilmark{1,2}, Hui Li\altaffilmark{1,2}, Weiqun Gan\altaffilmark{1,2}}
\email{yingbl@pmo.ac.cn;lfeng@pmo.ac.cn}

\altaffiltext{1}{Key Laboratory of Dark Matter and Space Astronomy, Purple Mountain Observatory, Chinese Academy of Sciences, 210032 Nanjing, China}
\altaffiltext{2}{School of Astronomy and Space Science, University of Science and Technology of China, Hefei, Anhui 230026, People's Republic of China}
\altaffiltext{3}{INAF-Turin Astrophysical Observatory, via Osservatorio 20, 10025 Pino Torinese (TO), Italy}
\altaffiltext{4}{School of Mathematics and Statistics, University of St. Andrews, North Haugh, St. Andrews, Fife, KY16 9SS, UK}

\begin{abstract}
The determination of the speed of Coronal Mass Ejections (CMEs) is usually done by tracking brighter features (such as the CME front and core) in visible light coronagraphic images and by deriving unidimensional profiles of the CME speed as a function of altitude or time. Nevertheless, CMEs are usually characterized by the presence of significant density inhomogeneities propagating outward with different radial and latitudinal projected speeds, resulting in a complex evolution eventually forming the Interplanetary CME. In this work, we demonstrate for the first time how coronagraphic image sequences can be analyzed with cross-correlation technique to derive 2D maps of the almost instantaneous plasma speed distribution within the body of CMEs. The technique is first tested with the analysis of synthetic data, and then applied to real observations. Results from this work allow to characterize the distribution and time evolution of kinetic energy inside CMEs, as well as the mechanical energy (combined with the kinetic and potential energy) partition between the core and front of the CME. In the future, CMEs will be observed by two channels (VL and UV Ly-$\alpha$) coronagraphs, such as Metis on-board ESA Solar Orbiter mission as well as Ly-$\alpha$ Solar Telescope (LST) on-board Chinese Advanced Space-based Solar Observatory (ASO-S) mission. Our results will help the analysis of these future observations, helping in particular to take into account the 2D distribution of Ly-$\alpha$ Doppler dimming effect.
\end{abstract}

\section{Introduction}
\bibliographystyle{apj}

Coronal Mass Ejections (CMEs) are large scale phenomena related with huge magnetic plasma explosions from the Sun. With high speeds even up to 3500 km s$^{-1}$ \citep{Gopalswamy2009}, these eruptions can release a large amount of energy (10$^{29}$-10$^{32}$ ergs) and plasma (10$^{15}$-10$^{16}$ g) ejected into the interplanetary space. When the accelerated plasma and the energetic particles propagate towards the Earth, geomagnetic storms can be caused. 

In recent decades, many different instruments have been designed to observe different layers of the Sun and to reveal different physical phenomena, such as space-based visible light (VL) coronagraphs able to monitor the lower corona 24 hours per day. The Large Angle Spectroscopic Coronagraph \citep[LASCO;][]{Brueckner1995} on-board the \emph{Solar and Heliospheric Observatory (SOHO)} mission successfully demonstrated that these coronagraphic instruments are capable to provide real-time information on the properties of solar eruptions, such as kinematics, masses, evolution of CMEs and other parameters. The COR1 and COR2 coronagraphs on-board the \emph{Solar TErrestrial RElations Observatory (STEREO)} mission \citep{Kaiser2008} have a big advantage to observe the propagation of CMEs simultaneously from two different perspectives, thanks to the STEREO twin spacecraft. These coronagraphs have offered the best way to investigate the geometrical structure of CMEs. Many methods to reconstruct CME's three-dimensional (3D) geometrical structure have been developed, such as mask fitting \citep{Feng2012a}, forward modeling, triangulation method, polarization ratio method \citep{Lu2017} and the graduated cylindrical shell (GCS) model \citep{Thernisien2006,Thernisien2009}. 

The UV Coronagraph Spectrometer \citep[UVCS; ][]{Kohl1995} on-board SOHO also allowed to investigate great details of CME events by using the UV emission lines (such as H I Ly-$\alpha$ 121.6 nm, O VI 103.2 and O VI 103.7 nm spectral lines), but its field of view (FOV) was limited to the spectrometer entrance slit (42 arcmin). Observations from UVCS have been used to derive plasma proton and electron temperatures, heavy ion kinetic temperatures, and elemental distributions \citep[e.g.][]{Akmal2001, Bemporad2007, Ciaravella2003,Ciaravella2006, Susino2016, Bemporad2018}, while all of these information cannot be obtained from VL coronagraphs alone. Unfortunately, after almost 17 years of successful observations of CMEs, solar wind, and other phenomena, the UVCS instrument has been no longer operational since 2012. 
A lot of previous studies based on the analysis of VL coronagraphic observations concentrate on the estimate of CMEs masses, and subsequently on deriving the kinetic and potential energies \citep[e.g.][]{Vourlidas2000,Colaninno2009, Carley2012, Feng2013}. Nevertheless, when researchers estimate the kinetic energy of a CME, they always use the unidimensional speed measured for the front or the core of the CME, and ignore its anisotropic kinematical properties.

Despite the importance of this topic, the 2D distribution of kinematical properties of CME plasma was studied only in a few previous works, and with many different techniques. \citet{Low1987} calculated the velocity field of a CME based on visual tracking of features in subsequent images. \citet{Tappin1999} calculated the outflow speeds of the solar wind by using a cross-correlation method, but they did not apply this method to CMEs. \citet{Robbrecht2004} used the modified Hough transform to filter the most significant signals of CMEs, and calculated their average speed while regarding the CMEs as bright ridges in (time, height) maps. They did not generate a velocity field over the whole CME. \citet{Colaninno2006} used optical flow methods to estimate the velocity fields of whole CMEs' bodies. \citet{Feng2015a} provided a method to investigate the CME kinematics via determining the radial mass transport process throughout the entire CME, which is capable of the estimation of the radial flow speed inside the CME. \citet{Braga2017} used the CORSET3D to determine the tie-points along the CME front automatically in three-dimensional (3D) space and derived the average two-dimensional (2D) speed of CMEs' fronts in the Stonyhurst coordinate system. Compared with these studies, all of these results have never shown the 2D kinetic and potential energy distributions of CMEs. 

In this paper, we successfully use the VL coronagraphic images to obtain the 2D speed map of the plasma within the body of a CME using a cross-correlation method, and to derive the first kinetic energy distribution of the CME. In the next few years, new coronagraphic instruments with new channels will be launched in space. The Metis instrument \citep{Antonucci2017} on-board ESA \emph{Solar Orbiter} mission (to be launched in 2020) will provide the first-ever simultaneous observations of the solar corona with FOV of 1.6-2.9$^{\circ}$ (the corresponding projected altitudes in solar radii from 1.7 to 3.1 R$_{\odot}$ when the spacecraft will be at closest approach at 0.28 AU), combining two different spectral bands: broad-band (580-640 nm) in the VL, and narrow-band in the UV emission from the neutral H I atoms (121.6 nm Ly-$\alpha$ line, the most intense line in the UV solar spectrum). Similar to Metis, the LST instrument \citep[Ly-$\alpha$ Solar Telescope;][]{Li2015} on-board the future Chinese \emph{Advanced Space-based Solar Observatory (ASO-S)} mission\citep[][to be launched in 2022]{Gan2015} will also observe the corona in the VL and in the UV H I Ly-$\alpha$ line, with FOVs from solar disk to 1.2 R$_{\odot}$ and 1.1 R$_{\odot}$ to 2.5 R$_{\odot}$. As we will discuss, CMEs' analysis technique being presented here will be very important to support future observations of solar eruptions that will be acquired by these instruments. 

The mechanism of formation of the H I Ly-$\alpha$ line from an optically thin corona is a combination of radiative and collisional excitations. The radiative component is generated by resonant scattering of the chromospheric and photospheric radiation by coronal neutral H I atoms, while the collisional component is due to the de-excitation of a coronal atom previously excited by collision with a free electron. When the CME moves outward from the sun, the intensity of H I Ly-$\alpha$ line due to radiative excitation will decrease due to the Doppler dimming effect. According to \citet{Noci1987}, the Doppler dimming factor is defined as follows:
 \begin{equation}
   \centering
   F_D (v_r)= {\int}_{0}^{+\infty} I_{ex}(\lambda-\delta\lambda)\Phi(\lambda-\lambda_0)d\lambda,
   \label{eq:doppler}
 \end{equation}
 where $I_{ex}(\lambda-\delta\lambda)$ is the intensity line profile of the chromospheric radiation. $\delta\lambda = \frac{\lambda_0}{c}v_r$ is the wavelength shift of the exciting profile due to the radial velocity of the moving plasma $v_r$, and $c$ is the light speed. $\Phi(\lambda-\lambda_0)$ is the normalized coronal absorption profile along the direction of the incident radiation, and $\lambda_0$ is the reference wavelength of the line transition. If the Doppler dimming factor can be estimated in advance, we could be able to obtain the electron temperature by the combination of the VL images and the UV H I Ly-$\alpha$ intensity observations. Without spectroscopic observations, our method to derive the velocity distribution of the CME can be considered as a useful diagnostic technique to estimate the 2D distribution of Doppler dimming factors of the observed CME in the future data. 
 
\citet{Susino2016} have first combined polarized VL images from LASCO/C2 and simultaneous data in the UV H I Ly-$\alpha$ line (1216 {\AA}) acquired by the UVCS spectrometer to estimate the CME plasma electron temperature, given the CME speed measured from LASCO images. The purpose of that work was to test the possible determination of CME plasma temperatures from the analysis of VL images and UV H I Ly-$\alpha$ intensities, without spectroscopic data. Subsequently, \citet{Bemporad2018} has rebuilt the synthetic images of total brightness (tB), polarized brightness (pB) and H I Ly-$\alpha$ intensities by using the output parameters from a numerical 3D MHD simulation of a CME. These synthetic images have been analyzed to measure the CME plasma temperatures, constrained by the simulated parameters, while taking into account Doppler dimming effect. In our paper, we calculate the velocity distribution of simulated CME \citep[on the basis of the method described in][]{Bemporad2018} and apply the same technique to a real event. 

In particular in this paper, we first estimate the measurement error of the speed obtained from our method by using the MHD simulated results in Section 2. Then we apply the cross-correlation method to a real CME event and obtain the first 2D speed and the Doppler dimming factor distribution of the CME in Section 3, and also the first kinetic energy distribution in 2D map. Moreover, we also derive the the partition and evolution of the mechanical energy, combined with the kinetic and potential energies, between the core and front of the CME.  Discussions and conclusions are described in Section 4.

\begin{figure*}[!th]
  \centering
  \includegraphics[width=1.\textwidth]{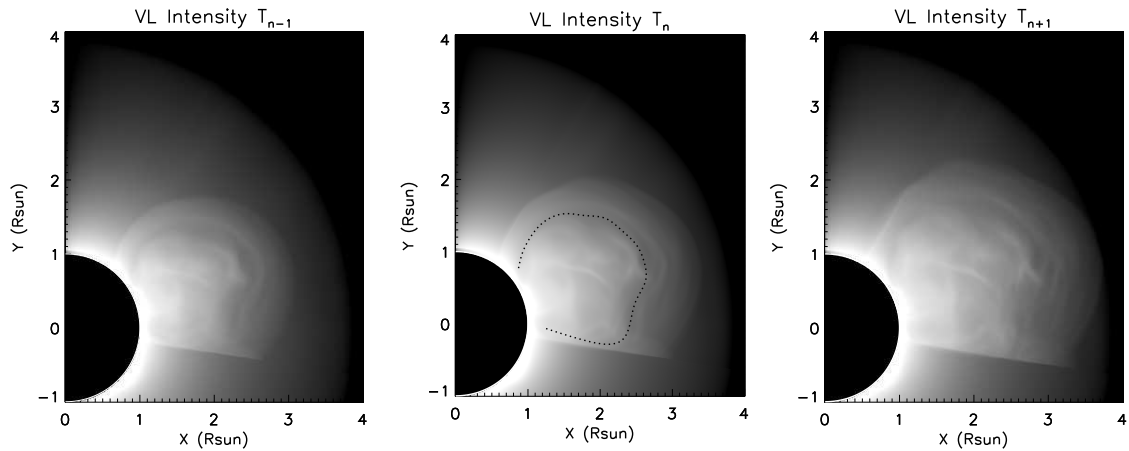}
  
  \caption{Resulting simulated tB intensities of a CME in coronagraphic images acquired in the VL channel. $T_{n-1}=23.2$ minutes (left panel), $T_{n}=26.1$ minutes (middle panel) and $T_{n+1}=29.0$ minutes (right panel). The time interval in the MHD simulation between single frames is 174 seconds. The dotted line represents the region of the CME core.}
  \label{fig:Model_VL}
\end{figure*}

\section{Measurement of the radial speed of the CME from synthetic VL images}

  

In the first part of this work we generate synthetic VL images of a CME using the same parameters from the MHD simulation of \citet{Pagano2014}. We refer to this work for details. The advantage to use VL synthetic images to test the technique for radial speed determination is that in these images the measured intensities are independent on the plasma outflow speed, different from what happens in the UV H I Ly-$\alpha$ images. Here we compute the visible-light tB from each simulated plasma element by integrating the contribution along a given line of sight (LOS) and then obtain the synthetic tB images. These have been derived from the computation of the Thomson scattered light from a single electron in each cell, multiplied by the electron number density of each cell, as the Thomson scattered light only depends on the position with respect to the solar disk and the scattering angle \citep{Billings1996}. 

\begin{figure*}[!th]
  \centering
  \includegraphics[width=0.9\textwidth]{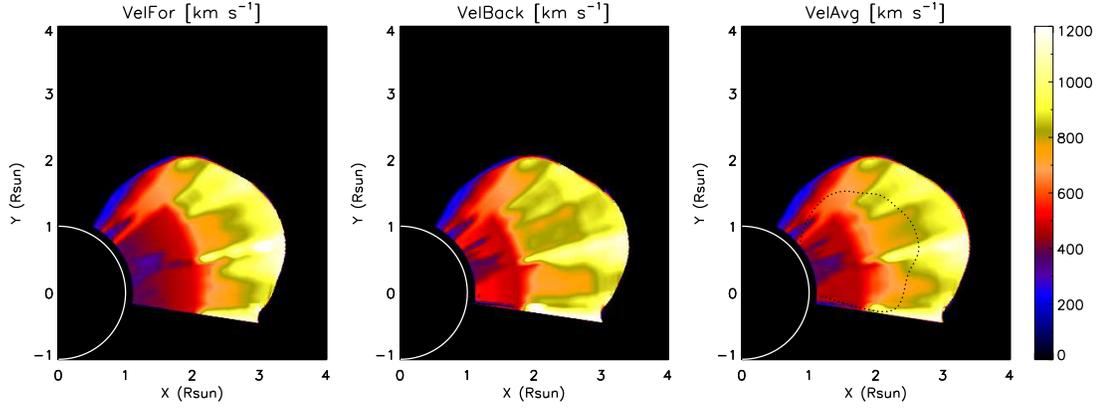}
  
  \caption{2D images of the measured radial velocities of the CME from the VL images by FS (left), BS (middle) and AS (right) methods. The dotted line represents the region of the CME core.}
  \label{fig:Model_measured_vel}
\end{figure*}

\begin{figure*}[!th]
  \centering
  \includegraphics[width=0.7\textwidth]{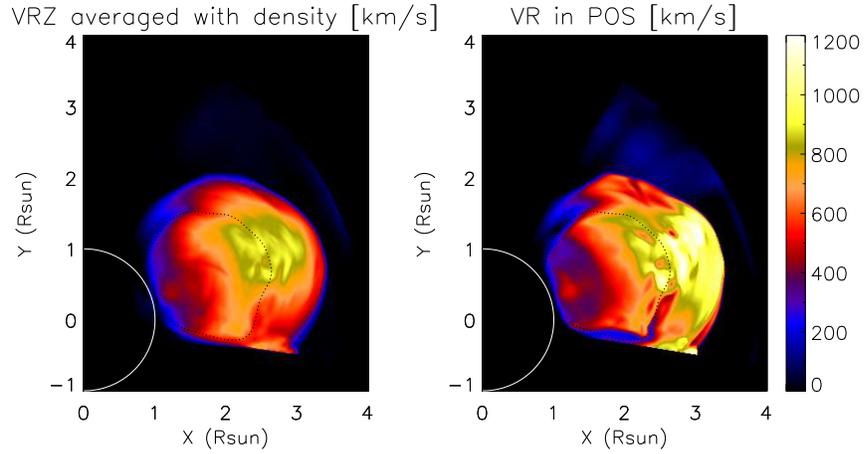}
  
  \caption{Two reference velocities obtained from the MHD simulation. Left: 2D image of the simulated radial velocity averaged with the electron number density (obtained by integrating along the LOS). Right: the cut of radial velocity in the POS. The simulation time is at $T=26.1$ minutes. The dotted line represents the region of the CME core.}
  \label{fig:Model_vel}
\end{figure*}

\begin{figure*}[!th]
  \centering
  \includegraphics[width=0.9\textwidth]{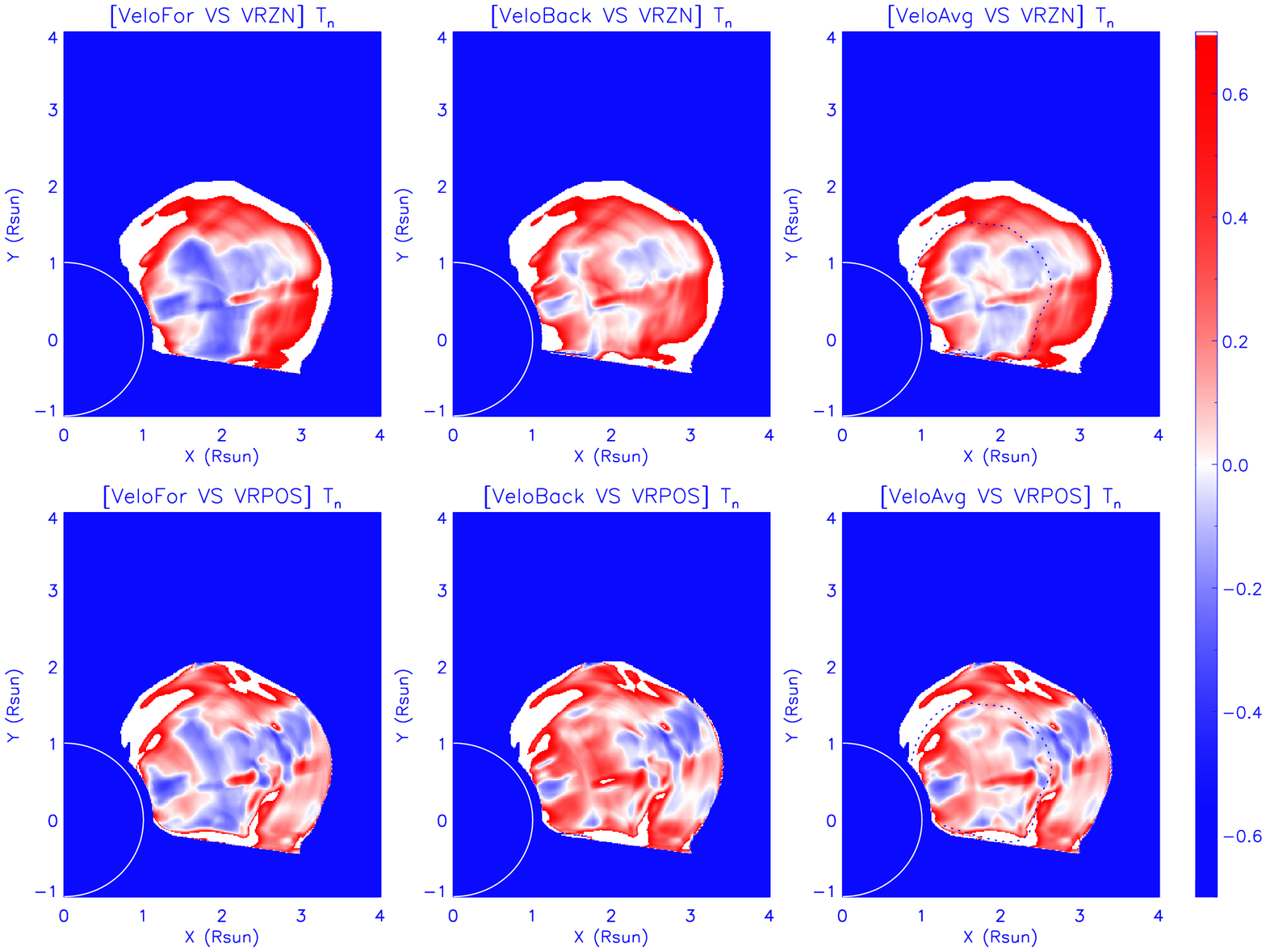}
  
  \caption{Top: relative comparisons between the measured radial velocities from the VL images and the simulated radial velocities averaged with density (FS, BS and AS, respectively). Bottom: relative comparison between the measured radial velocities and the simulated radial velocities in the POS (FS, BS and AS, respectively). The dotted lines represent the region of the CME core.}
  \label{fig:Model_vel_diff}
\end{figure*}

\subsection{Method}\label{sec:Method}

The resulting sequence of VL coronagraphic intensity images is shown in \autoref{fig:Model_VL}. These three synthetic VL images are at three simulated times $T_{n-1} =23.2$, $T_{n}=26.1$, $T_{n+1} = 29.0$ minutes, respectively.
We use these three VL images ( at $T_{n-1}$ , $T_n$, $T_{n+1}$ times) of the CME to measure its radial velocity at the $T_n$ time. Details of the method are listed: first, we use the normalized running-difference image (normalized by the previous image) to enhance the faint structures of the CME through a radial filter convolving with a Gaussian kernel (the SolarSoft routine {\tt filter\_image.pro}). Second, in order to measure the radial component of the plasma speed, we convert each VL filtered image from Cartesian to polar coordinates. This step allows us to extract the radial VL intensities at the three times at each fixed position angle, the angle going from 0 to 90 degrees and measured from the equatorial plane (corresponding to the East-West direction). Then, the radial intensity distributions at three different times are analyzed  in three steps: forward step (FS), backward step (BS) and average step(AS). In the FS, we determine pixel by pixel the radial shift by maximizing the cross-correlation between the signal in the actual frame (at $T_n$) extracted in a symmetric radial window centred on the considered point and the signal in a shifted radial window extracted in the next frame (at $T_{n+1}$). This first step is analogous to the method described in \citet{Bemporad2018}, but is improved here as we are going to describe. According to the assumption that the CME plasma is expanding faster in the front with respect to the core, we assume that the size of the shifted radial window increases linearly with the radial distance in the frame at $T_{n+1}$, so that we can obtain more corresponding signals, while the window at $T_n$ has constant size moving forward. Then the offset between the considered point and the central point of the matched window was regarded as the correspondent displacement. Subsequently, given the time interval between each frame (174 s), the displacement can be converted to the radial speed at the considered point. The BS analysis is similar, but with two differences with respect to the FS: 1) in the BS, we determine the maximal value of cross-correlation between the frame at the $T_{n}$ and the frame at the $T_{n-1}$; 2) the shifted radial window at $T_{n-1}$ is shrinking linearly, while the window size at $T_n$ moment is fixed. Finally, the average speed value obtained from FS and BS is considered as the AS. \autoref{fig:Model_measured_vel} shows the 2D radial velocity images as obtained with these three steps. From \autoref{fig:Model_measured_vel} it is clear that overall the speed maximizes at the CME front and decreases at lower heliocentric distances, as expected. Moreover, the plasma speed distribution within the CME body displays significant spatial inhomogeneities, due to the fact that different parts of the CME plasma are propagating with different speeds and accelerations. 

\subsection{Uncertainties in CME radial velocity}

In order to discuss results from our method, we first choose two input velocities ($t=26.1$ minutes) obtained from the MHD simulation as reference velocities. In fact, due to LOS integration, the result of any kind of analysis applied to 2D coronagraphic images will be a 2D velocity map, while the MHD simulation provides us with 3D data cubes. Here, following the analysis performed by \citet{Bemporad2018} for the optimization of plasma temperature determination in CMEs, we consider two reference velocities. The first one is simply the 2D cut of the radial velocity data cube in the POS. The second one is the average LOS velocity weighted over the electron number density $v_{LOS}={\int_{-\infty}^{+\infty} n_e v_r dz}/{\int_{-\infty}^{+\infty} n_e dz}$, considering that the VL brightness is proportional to the electron number density, $n_e$, of each cell, where $n_e=\rho/(0.83 m_p)$ and $m_p$ is the proton mass. Here, we assume that the typical coronal plasma condition of fully ionized atoms with He abundance equal to 10$\%$. These two different velocity distributions are shown in \autoref{fig:Model_vel} with the same color scale for comparison. It shows that significant differences between the two velocity distributions are present only around the nose of the CME front, where integration along the LOS are more significant. In the right panel of  \autoref{fig:Model_vel}, the speed distribution maximizes around the nose region, while it is not the case in the left panel.

The differences between the velocity maps measured from the synthetic VL images with the three methods (FS, BS and AS) and the two reference velocities mentioned above, subsequently normalized to the reference velocities, are so-called `relative comparisons', as shown in \autoref{fig:Model_vel_diff}. In particular, the top three panels in this Figure represent the relative comparisons between the measured radial velocities (from the FS, BS and AS, respectively) and the reference radial velocities averaged with the density, while the bottom panels show the relative comparisons between the measured velocities and the simulated velocities in the POS (FS, BS and AS, respectively). Considering in particular the comparison between the two AS maps and reference velocities (\autoref{fig:Model_vel_diff}, right column), it turns out that the measured radial velocities conforms better with respect to the reference speed weighted with the number density in the core of the CME (within 20$\%$ of uncertainty on average). On the other hand, the relative comparison with the POS speed (\autoref{fig:Model_vel_diff}, right bottom) shows a more uniform distribution, with a good velocity estimate (within 30$\%$ of uncertainty on average) of the plasma speed distribution in the POS all over the CME body, except for the flanks (where large errors up to about 100$\%$ are present). In general, we expect that the method described above provide us with a good uncertainty the radial velocity distribution within the CME body, with the exception of the regions located near the flanks of the eruption.

\begin{figure*}[!th]
  \centering
  \includegraphics[width=0.95\textwidth]{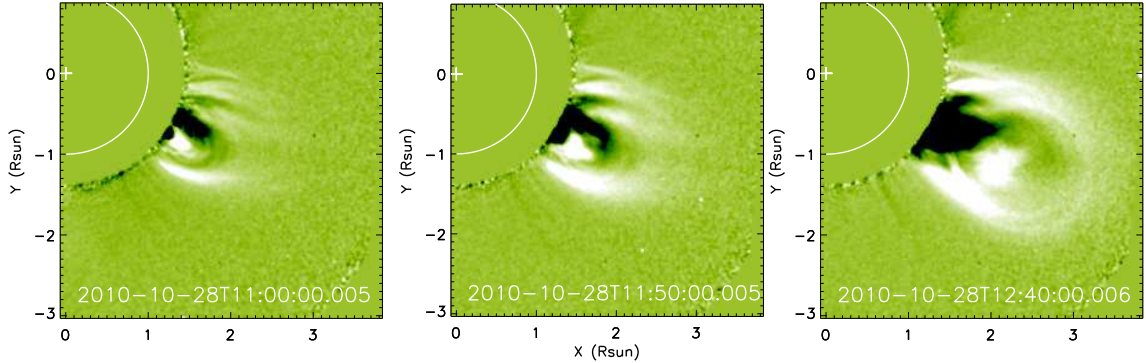}
  
  \caption{Examples of base-difference STEREO A/COR1 images acquired at three different times ($\sim$11:00, $\sim$11:50 and $\sim$12:40 UT) selected here with the purpose to show the CME early expansion. The pre-event background image acquired at 9:45 UT was subtracted. The white solid line and plus symbol denote the position of solar limb and center of the Sun, respectively, behind the COR1 occulter.}
  \label{fig:obs_base_img}
\end{figure*}

\begin{figure*}[!th]
  \centering
  \includegraphics[width=0.95\textwidth]{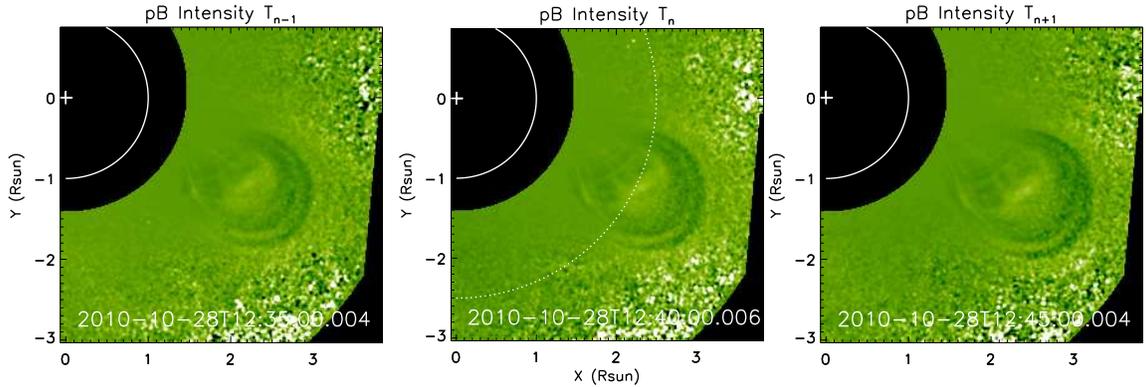}
  
  \caption{STEREO A/COR1 normalized running-difference images (acquired at$\sim$12:35, $\sim$12:40, and $\sim$12:45 UT). The position of a dotted line is at $r=2.5~R_{\odot}$. }
  \label{fig:obs_org_img}
\end{figure*}
\section{Analysis of a real event}
\subsection{Observations} \label{sec:observe}
To test the techniques described above on a real event, we selected a CME that was observed by STEREO coronagraphs. They have better time cadences with respect to SOHO coronagraphs. We decided to focus on a limb event in order to minimize possible projection effects due to plasma expansion out of the POS. On  2010 October 28, a CME with a classical three-part structure started to erupt at around $\sim$10:00 UT. The source region of this eruption was located in active region (AR) NOAA 11115, where a transient flare occurred. The CME, rising up from a streamer, was then observed both above the solar limb in the FOV of COR1 and COR2 coronagraphs on board the STEREO. COR1 images reach a lower heliocentric distance with a FOV going from 1.4 to 4 R$_{\odot}$, while COR2 FOV goes from 2.5 to 15 R$_{\odot}$. We fitted the Graduated Cylindrical Shell (GCS) model \citep{Thernisien2009} to the CME observed by COR1 A and B, and find that the propagation of the CME deviates from the POS with an angle of about 10 degrees as seen from STEREO A. Therefore, we assume that this limb CME propagates approximately in the POS of STEREO A.

In order to show the CME early expansion, three pB images of the CME acquired at three different times are shown in \autoref{fig:obs_base_img}. All images were processed by standard SolarSoft routines {\tt secchi\_prep.pro}. To better show the CME, these images are subtracted by the pre-event background image at around 09:45 UT, hence \autoref{fig:obs_base_img} shows the evolution of the CME excess brightness. The dark region in the core part of the CME is due to the presence of a streamer in the pre-event images. At the initial stage, the evolution of the CME shows only a slow change of local dynamics in the streamer. Then, with the acceleration phase, the CME structure separates from the streamer, evolving into a typical bubble shape. 
\begin{figure*}[!th]
  \centering
  \includegraphics[width=1.\textwidth]{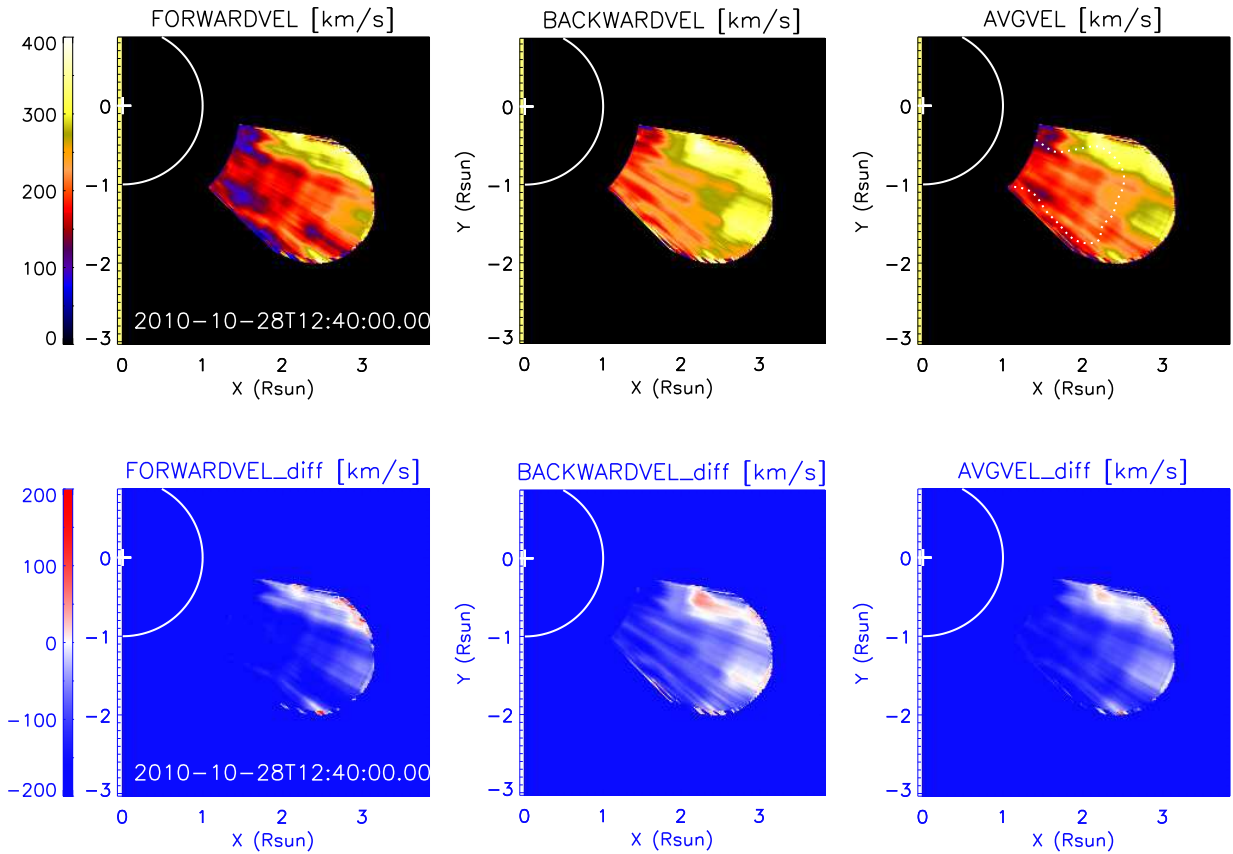}
  
  \caption{Top: 2D velocity maps measured from the VL images (FS, BS and AS). Bottom: 2D maps showing the differences between the 2D radial speed map measured from the analysis of VL images in the CME body and the speed value measured at the CME nose. The dotted line represents the region of the CME core defined in \autoref{sec:kinematic} and shown in \autoref{fig:min_base_img}.}
  \label{fig:vel_diff_img}
\end{figure*}

\begin{figure*}[!th]
  \centering
\end{figure*}

\begin{figure*}[!th]
  \centering
  \includegraphics[width=0.5\textwidth]{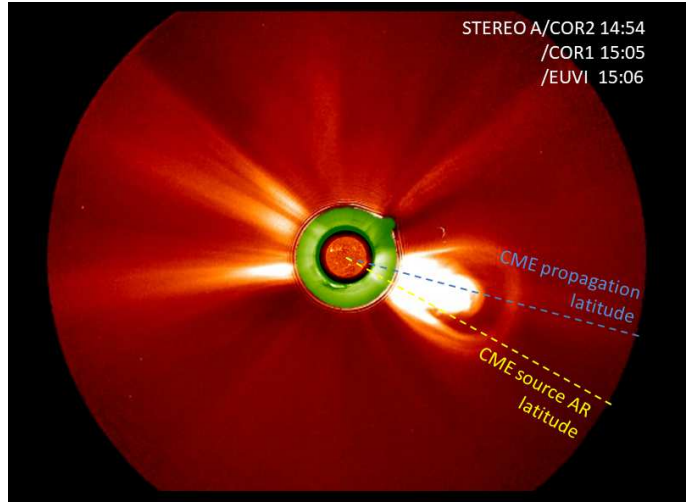}
  
  \caption{Superposition between STEREO A/EUVI, COR1 and COR2 images acquired on October 28, 2010 at 15:06, 15:05 and 14:54 UT, respectively. This Figure shows the difference between the source AR latitude at 29$^\circ$S (yellow dashed line) and the CME propagation latitude around $\sim 15^\circ$S (cyan dashed line).}
  \label{fig:cme_latitude}
\end{figure*}
\subsection{2D maps of the CME velocities}

To test the velocity measurement technique described in \autoref{sec:Method}, we selected here only three subsequent images observed from STEREO A/COR1 at three different times. The advantage of the STEREO COR1 data (compared with other coronagraphs) is that its high temporal resolution of 5 minutes is close to the time interval of the simulated VL image. We use the normalized running-difference images and subsequently enhance the contrast of the CME structure to measure the radial velocities of the CME in each pixel. As shown in \autoref{fig:obs_org_img}, many detail structures in the CME body can be observed in the resulting images. We mark these three sequential images (acquired at $\sim$12:35, $\sim$12:40, $\sim$12:45 UT) as $T_{n-1}$, $T_{n}$ and $T_{n+1}$ times, respectively. Through three steps (FS, BS and AS) in our method, the radial velocity maps have been derived, which are shown in \autoref{fig:vel_diff_img} (top row). It turns out that around 12:40 UT the speed of the CME body is not larger than 400 km s$^{-1}$, and the front speed is higher than the core, as expected. Moreover, the radial speed also maximizes around the northward flank of the CME, suggesting a significant latitudinal asymmetry in the CME propagation. The interpretation of this asymmetry is shown in \autoref{fig:cme_latitude}: while the source AR was located at a latitude of 29$^\circ$S, the nose of the CME front (representing the most likely CME propagation direction) was centered around $\sim 15^\circ$S. The CME was deflected northward with respect to the source AR. Comparing the shape of the CME front between that observed at 12:40 UT (\autoref{fig:obs_base_img}, right panel) and at 14:54 UT (\autoref{fig:cme_latitude}), it is almost circular at 12:40 UT, while that is not the case in the FOV of COR2, the front in the northward part of the CME seems moved further, this is well reproduced by the faster radial speed we derived in the 2D velocity distribution.

Given the 2D velocity maps, asymmetries and inhomogeneities in the 2D velocity distribution can be better explored by comparing these 2D velocities with the front speed measured along the propagation direction of the CME nose. In fact, this latter quantity is usually assumed in the literature to be representative of the speed in the whole CME body. The CME nose speed here was derived by extracting a radial slice in the excess brightness images to track its propagation along the nose direction; the resulting nose speed is about $\sim$320 km s$^{-1}$. The bottom panels of \autoref{fig:vel_diff_img} show the differences between the 2D radial speed (\autoref{fig:vel_diff_img}, top row) measured from the VL images in the CME body and the speed value measured at the CME nose. The comparison shows again that in some parts of the northward flank of the CME the radial plasma speed is even higher than the nose speed. The reasons of that are as follows: 1) as mentioned above, the CME was propagating northward with higher radial speed, hence a latitudinal north-south asymmetry was expected, and 2) as we showed with the analysis of synthetic data the uncertainties in the estimate of the CME velocity maximize at the flanks of the eruption. Moreover, as previously discussed by \citet{Bemporad2018}, each packet of plasma in the 2D image, in principle, could have a different component of velocity along the LOS. However, this component cannot be measured through the analysis of coronagraphic images, and can be estimated only with spectroscopic data (from the Doppler shift of spectral lines), as it was done many times in the literature with the analysis of SOHO/UVCS observations of CMEs \citep[see e.g.][]{Lee2009,Raymond2003}.

As we are going to show, the 2D maps of CME radial velocity derived in the previous Sections can be used to estimate the 2D distribution of further parameters that can be of interest. First of all, the velocity map can be employed to derive the 2D distribution of Doppler dimming factors for the UV H I Ly-$\alpha$ intensity, a parameter that will be very important for the future analysis of CME coronagraphic observations acquired in this band (see \autoref{sec:dopplerdim}). Moreover, these velocity maps can be used to investigate the 2D distribution and time evolution of the plasma kinetic energy in the CME body, that can provide interesting information on the energetic of the CME (see \autoref{sec:kinematic}).


\begin{figure*}[!th]
  \centering
  \includegraphics[width=1.\textwidth]{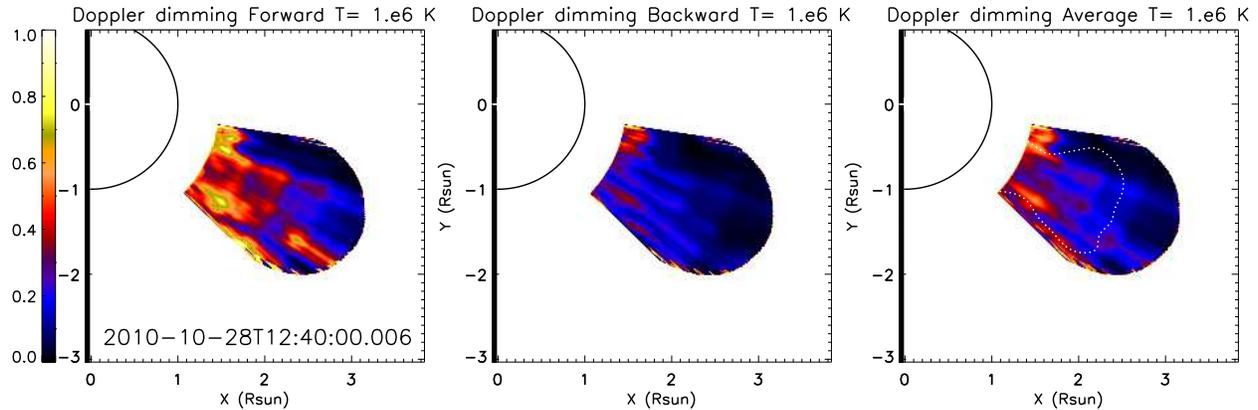}
  \caption{The distribution maps of the Doppler dimming factors (FS, BS and AS). The dotted line represents the region of the CME core, shown in  \autoref{fig:min_base_img}.}
  \label{fig:dopplerdim}
\end{figure*}

\subsection{Doppler dimming factor} \label{sec:dopplerdim}
In the UV spectral range, the primary physical processes responsible for atomic excitation (followed by spontaneous emission) are the collisional excitation and the photo-excitation. These are also the main contributions of the H I Ly-$\alpha$ emission in the optically thin corona, while other mechanisms could be neglected, such as the Thomson scattering of chromospheric radiation by free coronal electrons. The observations from the coronagraphs only provide 2D images, according to \citet{Susino2016}, the radiative component of the H I Ly-$\alpha$ can be expressed as
\begin{equation}
  \centering
  N_{e}\equiv \int_{LOS}n_edz \approx<n_e>L
\end{equation}
\begin{equation}
  \centering
  I_{rad}\propto R(T_e) \cdot F_D(v_r) <n_e>L,
  \label{eq:radinten}
\end{equation}
where $N_e$ is the column density which can be obtained from VL images, $<n_e>$ is the average electron density of the CME plasma along the LOS, and $L$ is the thickness of the CME plasma along the LOS. If the thickness $L$ of emitting plasma is known, the column density $N_e$ can be converted to volume density $<n_e>$ in units of cm$^{-3}$. In \autoref{eq:radinten}, $R(T_e)$ is the element ionization fraction, which is a function of the electron temperature $T_e$, and $F_D(v_r)$ is the Doppler dimming factor, as defined in \autoref{eq:doppler}. In the future, if the Doppler dimming factor and the thickness of the plasma along LOS can be determined, combining the observations of VL and UV intensities, we could estimate the temperature distribution of the CME plasma, according to \autoref{eq:radinten}. For instance with a similar analysis (based only on VL and UV H I Ly-$\alpha$ intensities) \citet{Susino2016} recently found that CME cores are usually associated with cooler plasma (T $\sim10^6$ K), and that a significant increase of the electron temperatures is observed from the core to the front of the CME (where T $>10^{6.3}$ K). 

However, in the analysis described above the Doppler dimming factor will be affected not only by the radial velocity of the CME plasma, but also by the shape of normalized coronal absorption profile, $\Phi(\lambda)$, and by the intensity spectrum of incident profile, $I_{ex}$, as described by \autoref{eq:doppler}. The normalized coronal absorption profile, here, has been assumed to be Gaussian with a width $\sigma_{\lambda}$, under the assumption of a Maxwellian velocity distribution of the absorbing particles, where 
\begin{equation}
  \centering
  \sigma_{\lambda}=\frac{\lambda_0}{c}\sqrt{\frac{k_BT_k}{m_p}}   ~~~({\rm cm}).
  \label{eq:Gauss_width}
\end{equation}
In \autoref{eq:Gauss_width}, $k_B$ is the Boltzmann constant, and $T_k$ is the kinetic temperature. To calculate the $F_D$, we assumed here an averaged value of the exciting radiation coming from the solar disk, and adopted the SOHO/SUMER chromospheric spectral profile observed at solar minimum in July 1996 \citep{Lemaire2002}. The kinetic temperature is assumed $T_k=10^6$ K for the normalized coronal absorption profile. The resulting 2D distribution of the Doppler dimming factors is shown in \autoref{fig:dopplerdim}. Due to the low speed of the CME body, the distribution of the $F_D$ factors reflect the anisotropies in the 2D distribution of CME velocity, while $F_D$ decreases sharply in the pixels where the CME plasma is expanding faster than $\sim 300$ km s$^{-1}$. Resulting 2D distribution of the $F_D$ factors further confirms that \citep[as recently shown by][]{Bemporad2018} the UV intensity in the CME front will be severely attenuated due to the larger plasma radial speed, as can be also revealed by comparing the distribution of the Doppler dimming factors in \autoref{fig:dopplerdim} with the radial speed distribution of the CME in \autoref{fig:vel_diff_img}. In the future, when we will estimate the electron temperature of CMEs propagating with a very large speed, it will even possible to neglect the contribution of the radiative component, and to consider only the collisional component of H I Ly-$\alpha$ intensity.
\begin{figure*}[!th]
  \centering
  \includegraphics[width=0.63\textwidth]{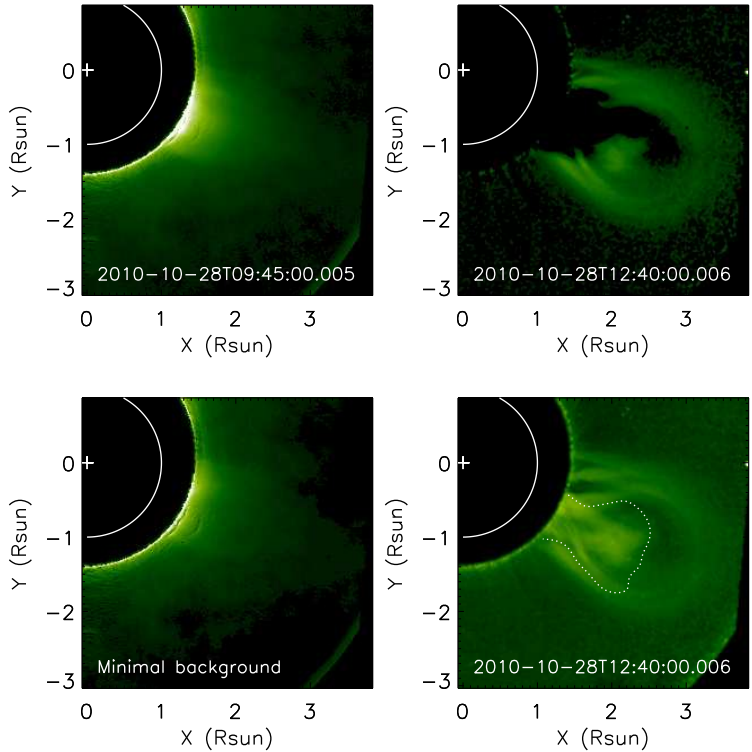}
  \caption{Left column: comparison between two different background pre-event images. The upper one is the pre-event background image acquired at $\sim$09:45 UT, while the lower one is the 24 hr minimum background image. Right column: base-difference images at $\sim$12:40 UT obtained by subtracting the two different background images shown in the left column. The dotted line represents the region of the CME core, which is defined in \autoref{sec:kinematic}.}
  \label{fig:min_base_img}
\end{figure*}

\begin{figure*}[!th]
  \centering
  \includegraphics[width=1.\textwidth]{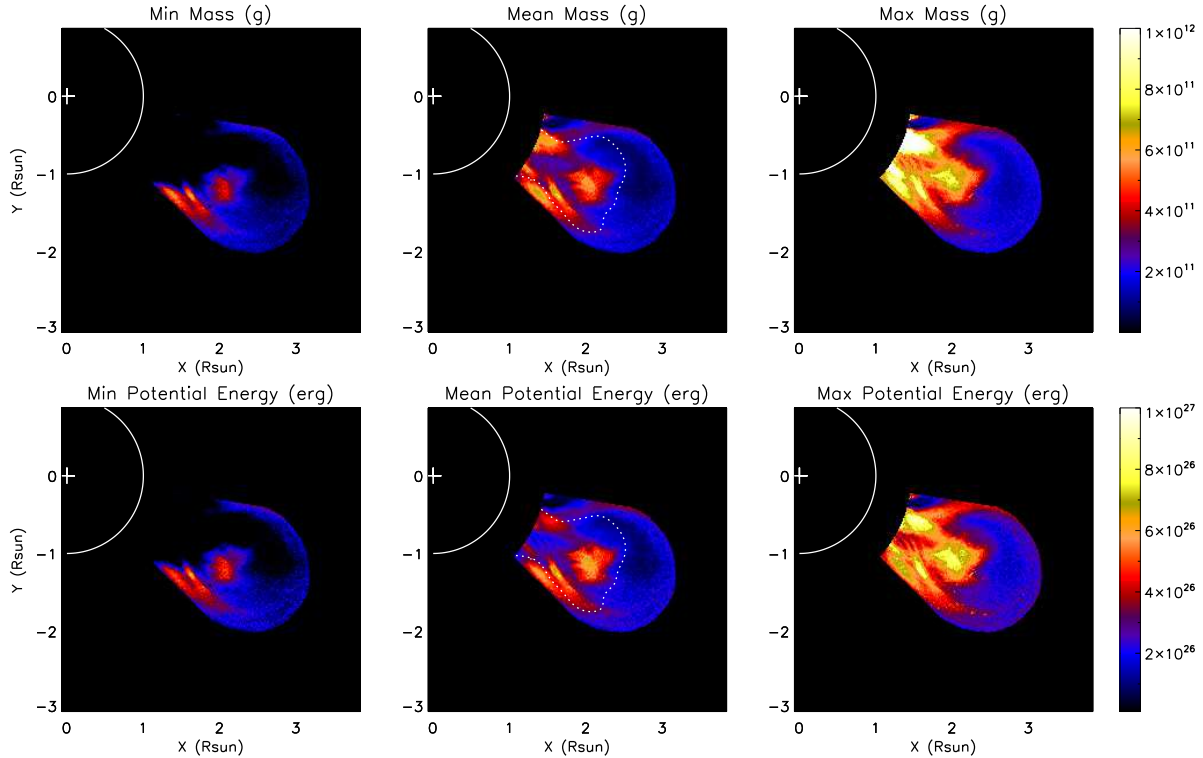}
  \caption{Top: The mass distribution images of the CME obtained by subtracting the pre-event image, the average between the pre-event and the 24 hr minimum image as well as the 24 hr minimum image, respectively. Bottom: The potential energy distributions of the CME obtained by subtracting to the same image the different background images shown in the top panels. The dotted lines represent the region of the CME core.}
  \label{fig:potential_img}
\end{figure*}

\subsection{Energies of the CME} \label{sec:kinematic}

To estimate the CME mechanical energy (including the potential and the kinetic energies), the first quantity that we need to estimate is the CME mass. There are different ways to achieve this, e.g., by using EUV dimming, spectroscopic observations, and the most common way, from Thomson scattering using VL coronagraphic images \citep[e.g.][]{Aschwanden2009, Aschwanden2016, Tian2012, Colaninno2009, Carley2012, Feng2013}. In this work, the Thomson-scattering approach is applied and the STEREO/COR1 pB images were used. Many previous authors \citep{Mierla2011, Howard2015a, Howard2015b} have shown that in COR1 observations of CMEs the Thomson-scattered signal will be strongly contaminated by H$\alpha$ emission from the prominence, subsequently resulting in the overestimation of the CME mass. Here, we emphasize that in our event there is no filament structure that erupted and propagated into the FOV of COR1 instrument, thus the contribution of H$\alpha$ emission can be ignored and the mass measured from COR1 can be considered as reliable.

The major part of the signal of white-light coronagraph images, yielded by Thomson scattering, is linked linearly to the density of the scattering free electrons. The major non-Thomson scattered contributions, like stray light and dust scattering (the F-corona), are usually removed by subtracting the pre-event background image. The remaining image brightness then should contain the CME intensity distribution produced by Thomson scattering alone by electron in the CME plasma. In our event, however, before the appearance of the CME, there exists a dense streamer for a long time, and subtracting the pre-event background image (\autoref{fig:min_base_img}, top left), which also includes the streamer structure, will result in negative values appearing in many pixels of the difference image (\autoref{fig:min_base_img}, top right), causing a large underestimate of the CME mass. Therefore, here we consider the image resulting from the subtraction of the pre-event image as the lower limit estimate of the CME mass. Alternatively, a different approach is also possible by subtracting the 24 hr minimum background image (\autoref{fig:min_base_img}, bottom left), in which each pixel has the minimal brightness observed from $\sim$00:00 UT to $\sim$23:55 UT; in this way we obtained the new difference image shown in \autoref{fig:min_base_img} (bottom right). This image containing some streamer structure can be then converted to the mass image and then be regarded as the upper limit of the CME mass. The mass distributions of the CME, derived from the pB images via the SolarSoft routine {\tt eltheory.pro}, are presented in \autoref{fig:potential_img}. Here, we derived two different mass distributions by subtracting the two different background images mentioned above. The region of interest we consider to estimate the CME mass corresponds to the region where we measured the radial velocity, while the remaining part is ignored here. Then we can calculate the potential energy distribution of the CME, hence the energy required to lift the CME plasma from the solar surface to its position $R$, according to the equation
\begin{equation}
  \centering
  E_p=\int_{R_{\odot}}^{R}\frac{G M_{\odot} m_{\rm CME}}{r^2}dr~~~({\rm erg}),
  \label{eq:potential}
\end{equation}
where $G$ is Gravitational constant, $M_{\odot}$ is the solar mass, $m_{CME}$ is the mass of the CME plasma, and $R$ is the radial distance from the solar centre. The potential energy of the CME at $\sim$12:40 UT is shown in \autoref{fig:potential_img} (bottom panels). 

\begin{figure*}[!th]
  \centering
  \includegraphics[width=1.\textwidth]{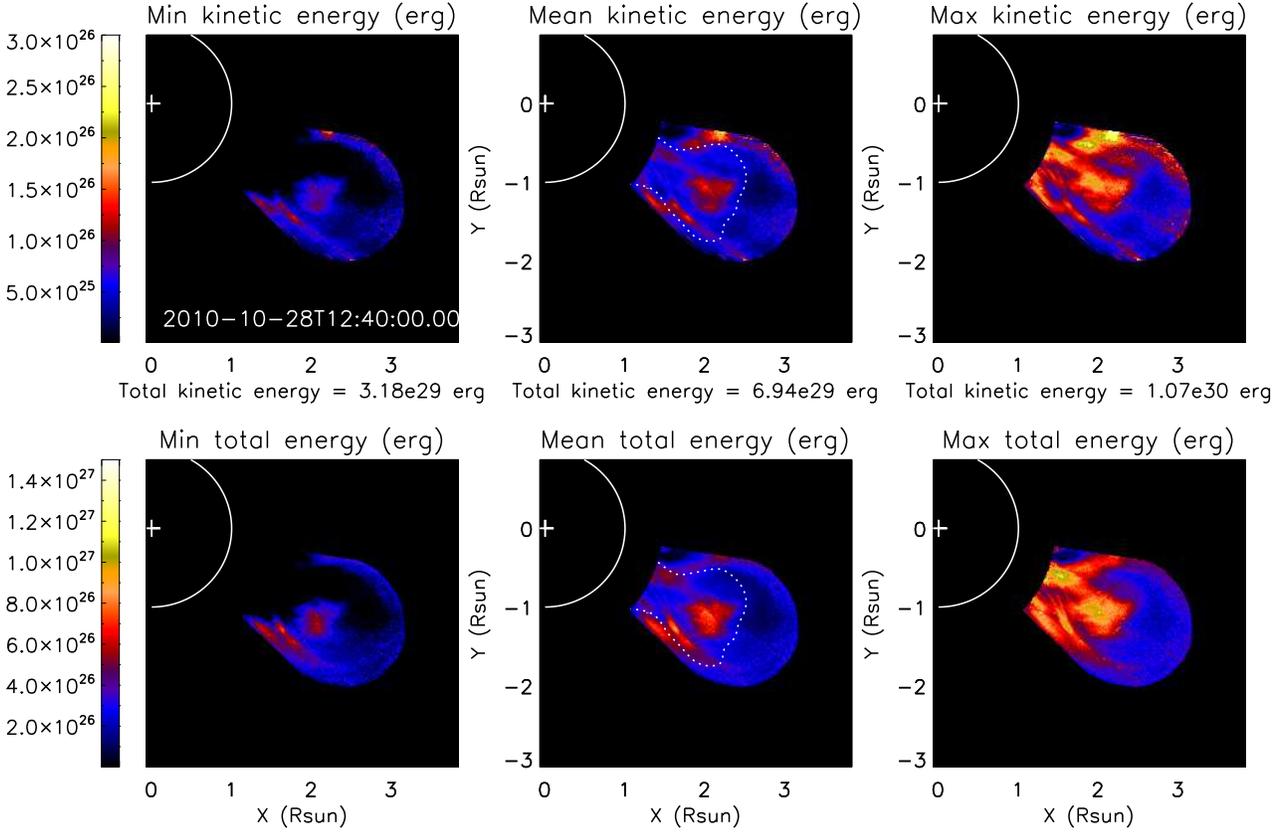}
  \caption{Top: 2D images of the kinetic energy of the CME at 12:40 UT, using the average speed (obtained from the AS method) with the different background subtractions. Bottom: 2D images of the CME total mechanical (kinetic plus potential) energy at 12:40 UT. The dotted lines represent the region of the CME core.}
  \label{fig:kinetic_img}
\end{figure*}
On the other hand, the CME kinetic energy distribution is given by
\begin{equation}
  \centering
  E_k=\frac{1}{2}~m_{\rm CME}~v^2~~~({\rm erg}),
  \label{eq:kinetic}
\end{equation}
where, by indicating with $v_{\theta}$ the plasma velocity along the latitudinal direction in the POS, and with $v_{\phi}$ the plasma velocity along the longitudinal direction, the total velocity $v$ is given by $v=\sqrt{v_{r}^2+v_{\theta}^2+v_{\phi}^2}$. As we mentioned in \autoref{sec:observe}, this CME was almost propagating in the POS, so the measured radial velocity in the POS can be used as an approximate to its 3D radial speed $v_{r}$, in space. Unfortunately, we cannot estimate the latitudinal and longitudinal velocities directly from our method. Here, we choose two ways to estimate the total kinetic energy of the CME: 1) by only considering the radial velocity; 2) by measuring the tangential speed of the CME plasma with latitudinal slices in the VL images (such as the dotted line shown in \autoref{fig:obs_org_img}), and by assuming that the longitudinal velocity of the CME plasma is equal to the measured latitudinal velocity. By using the polar coordinate data mentioned above, it is easy to make slices along latitude at the same radial altitude; in particular we use a width of the slices of 0.125 R$_{\odot}$. For example, to analyze this CME at 12:40 UT, we make around 20 latitudinal slices and obtain time-distance plots at different positions. Subsequently, by comparing all velocity results, we acquire the maximal latitudinal speed of the CME flank, and regard it as the average speed of a packet of plasma in each pixel cell, by assuming an equality between the longitudinal and latitudinal speeds. The kinetic energy will be overestimated when the latitudinal speed distribution should be inhomogeneous and small.

Finally, the derived kinetic energy distributions of the CME body are shown in \autoref{fig:kinetic_img}. Here, we choose the average velocity obtained from the AS procedure, multiplying with the different mass distributions of the CME shown in \autoref{fig:potential_img}, to estimate the kinetic energy and mechanical energy of the CME at 12:40 UT. The total kinetic energy is of $\sim$(3.18 - 10.7) $\times$ 10$^{29}$ erg, and the average of the total kinetic energy is of 6.94 $\times$ 10$^{29}$ erg. Actually, when we just calculate the kinetic energy by the radial velocity, the result is only 7$\%$ less than the kinetic energy with all the three components of velocity taken into account.

\begin{figure*}[!th]
  \centering
  \includegraphics[width=0.5\textwidth]{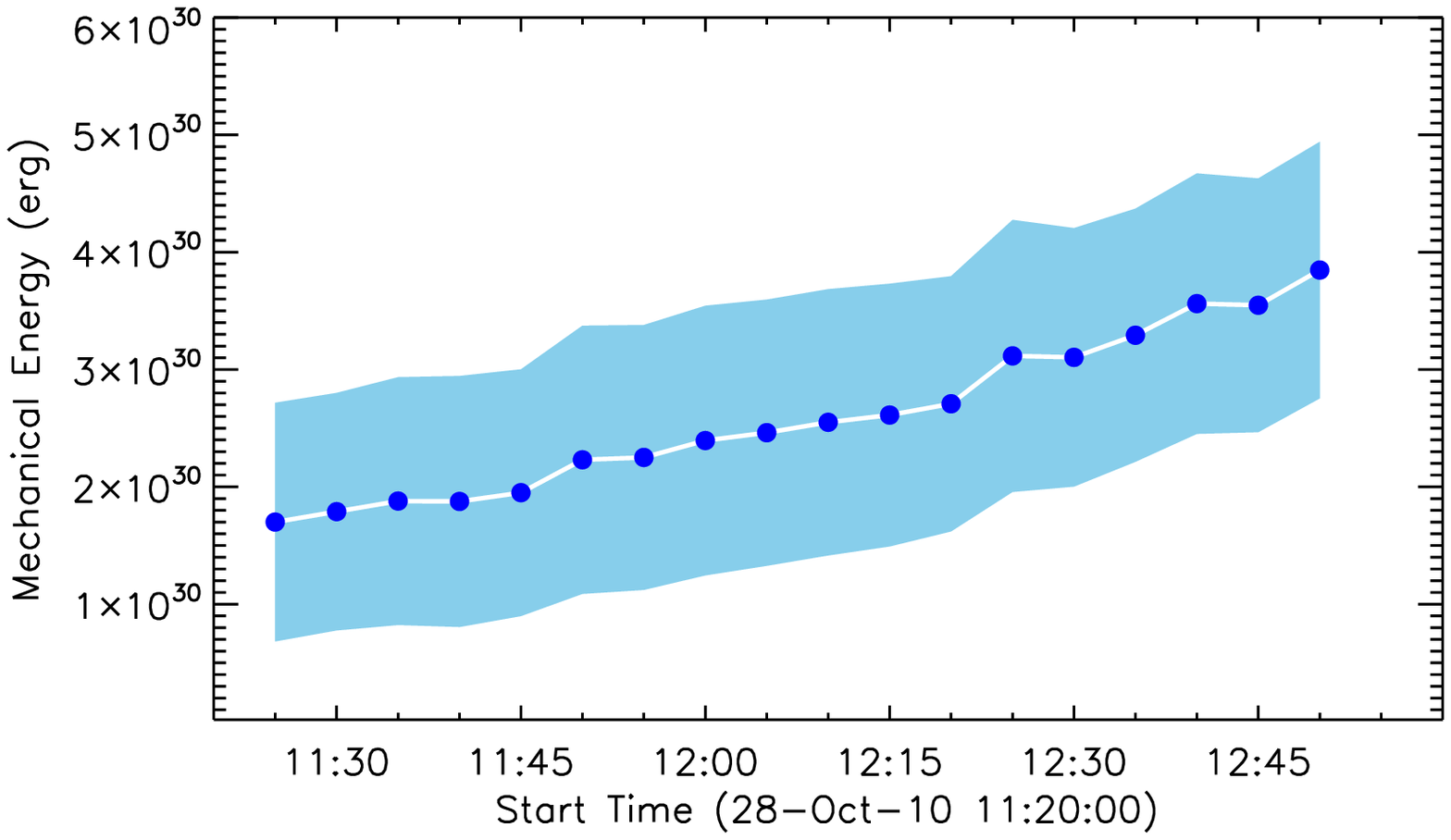}
  \includegraphics[width=0.5\textwidth]{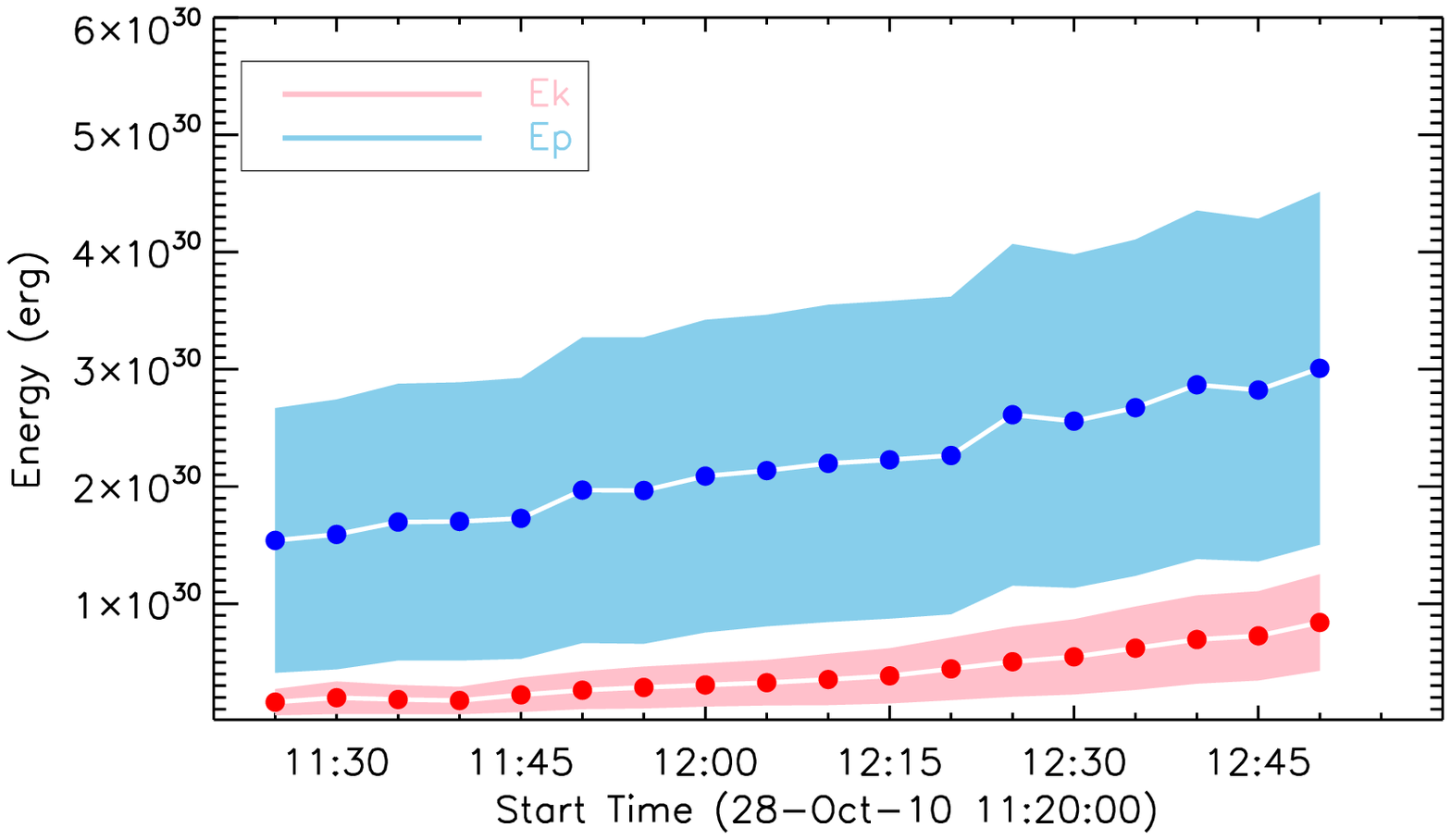}
  \caption{Top: Evolution of the mechanical energy of the CME with the time. Bottom: Evolution of the kinetic and potential energy of the CME with time. Pink (blue) region is the range of the kinetic (potential) energy. Red (blue) dots show the average value of the kinetic (potential) energy. }
  \label{fig:Machan_img}
\end{figure*}

We obtain the CME's mechanical energy, $E_M$, combining its potential and total kinetic energy. The evolution of total $E_M$, $E_k$ and $E_p$ from 11:25~UT to 12:50~UT is shown in \autoref{fig:Machan_img}. Notice that during this period the structure of the CME, still in the FOV of STEREO A/COR1, can be distinguished and be used to measure its velocities with our method. In \autoref{fig:Machan_img} (top panel), the total mechanical energy of the CME enhances around from 1.7 $\times 10^{30}$ to 3.9 $\times 10^{30}$ erg with a stable energy increasing rate of 4.3  $\times 10^{26}$ erg s$^{-1}$, the blue dots represent the average mechanical energy, while the blue shadow indicates uncertainties propagated from the mass measurements. For this CME, due to its low speed movement, the kinetic energy is far smaller than its potential energy, as shown in  \autoref{fig:Machan_img} (bottom panel). The increase of the kinetic energy is due to both the mass and the speed increase. The dots and shadow regions also represent the average value and error of potential and kinetic energy in blue and red (pink) color, respectively. \citet{Vourlidas2000} analyzed the energy budget of 11 CMEs and found that the kinetic energy is smaller than the potential energy for relatively small CMEs, but larger for fast CMEs ($v_{cme} \ge$ 600 km s$^{-1}$). Due to the conservation of the total energy (consisting of the kinetic, potential and magnetic energies, and neglecting other energy forms such as thermal energy and wave energy), the author thought that this result indicates that slow CMEs are magnetically driven, a conclusion which is reinforced by the small magnitude of the plasma $\beta$ parameter.


It is well-known that a typical CME has a three-part structure, consisting of (1) a bright leading edge, (2) a dark cavity, and (3) a bright dense core \citep{Illing1985}, originated from the explosive magnetic flux rope system. Many authors believe that the dark void and the bright core correspond to the whole flux rope and the magnetic dips of the flux rope \citep[e.g.,][]{Low1995, chen1996, Gibson2006, Cheng2012}, respctively. \citet{Cheng2012} has analysed the density and temperature of the CME leading front and  revealed that the CME bright front is resulted from the compression of ambient plasma when the CME expands its core region. Thus, in our work, we also discuss the energy distribution of the CME's core and front. In order to define these two ``core" and ``front" regions in a more quantitative way, we follow this procedure: for each VL base-difference image subtracting the 24 hr minimum background, we define ``by hand" a region of interest where the core should be located, taking this region as big as possible. Then, in this region we identify the maximum intensity $I_{max}$ and define as pixels belonging to the CME core those with an intensity being between 1 and a fraction $f$ of $I_{max}$ (optimizing the value of $f$, here $f=0.07$). Then, once we identified all the pixels belonging to the core in this way, we assume that all the rest of the CME region belong to the ``front$+$void" instead. In this way we make a separation between the CME ``core" on one hand, and the CME ``front$+$void" on the other hand, as done by \citet{Howard2015a,Howard2015b}. The CME core is marked by the dotted line in \autoref{fig:min_base_img} (bottom right).

\begin{figure*}[!th]
  \centering
  \includegraphics[width=0.5\textwidth]{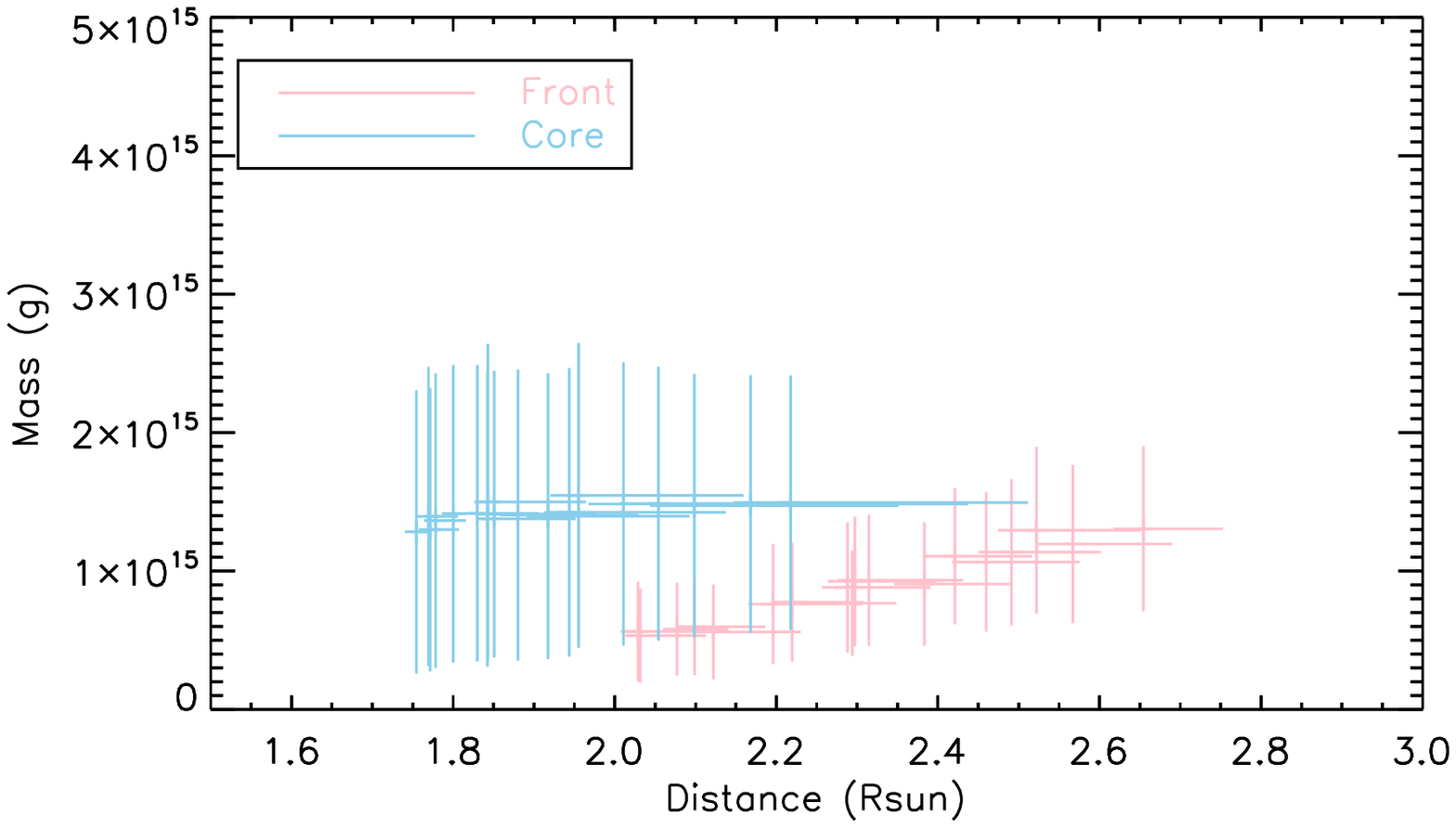}
  \includegraphics[width=0.5\textwidth]{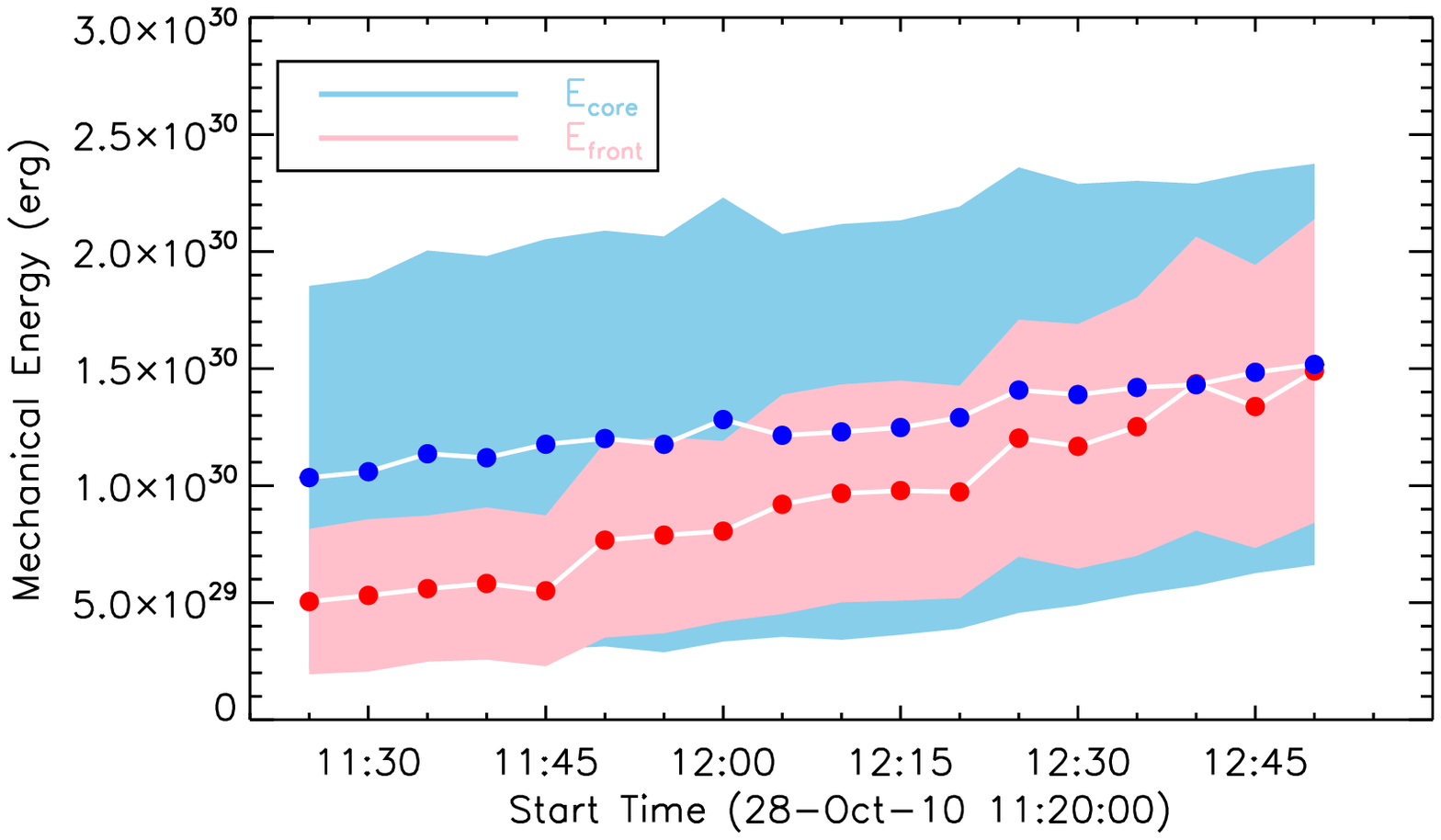}
  \caption{Top: Mass vs. centroid distance measurements of the core (blue) and its front (pink) components of the CME. Error bars are estimated by subtracting the different backgrounds. Bottom: Evolution of the mechanical energy measured from the CME core (blue shadow) and front (pink shadow) parts. Blue and red dots represent the average mechanical energy of the CME core and front, respectively. The shadows show the error ranges of the mechanical energy.}
  \label{fig:Machan_part_img}
\end{figure*}

The relationship between the mass of the CME and the calculated centroid distance is shown in \autoref{fig:Machan_part_img} (top panel). In \autoref{fig:Machan_part_img} the blue signs indicate the mass evolution of the CME core, which is almost constant with the centroid distance (or with the time), while the mass of the CME front, marked by the pink symbols, increases with its propagation toward the interplanetary space. The centroid distance is calculated by $r_c=\frac{\sum m_i r_i}{\sum m_i}$, where $m_i$ and $r_i$ denote the mass and distance from Sun center, respectively, of each pixel. The uncertainties shown here are propagated from the CME mass measurements with two different background subtractions. 
This result shows that during the CME propagation a significant mass is being added mostly at the CME front and not at the core, and this is in agreement with the fact that the core should be more ``magnetically isolated" from the rest of the CME because it is surrounded by the flux rope, while the front, on the one hand, being made also by coronal plasma dragged up during the CME propagation is expected to have a mass increase, which also has been reported by many authours \citep{Vourlidas2000,Cheng2012,Howard2015b}. The same result was already demonstrated with the analysis of SOHO/UVCS data deriving the elemental distribution of CME plasma embedded in the front compared with the plasma in the surrounding corona \citep{Ciaravella2003}. But on the other hand, \citet{Feng2015c} used ``snow-plow'' model to calculate the mass contribution of the pile-up solar wind and found that the effect of the solar wind pileup is not sufficient to explain the observed mass increase. Many authors think the increase of the CME mass is mainly due to the mass supplement by the outflow from dimming regions in the low corona \citep{Aschwanden2009, Jin2009, Tian2012}.  What's more, \citet{Howard2018} studied the evolution of the volume electron density of 13 CMEs' fronts and found that there is no obvious pile-up observed in the front region of the CMEs from 2.6-30 R$_{\odot}$. Whether the mass increase outside the core region of CME mostly comes from the solar wind pileup or the dimming region outflow requires further analysis, and is beyond the scope of this paper.

In \autoref{fig:Machan_part_img} (bottom panel), red and blue dots represent the average mechanical energy of the CME core and front, respectively, while the shadows indicate the upper and lower limits of the mechanical energy due to the mass measurement with different background subtractions. This Figure shows that at the initial stage of the CME the mechanical energy of the core is higher than that of the front. Then, with the propagation of the CME, the energy of the front rises more rapidly, reaching up that of the core. This is likely an indirect consequence of the CME formation: at the early stages the eruption is driven by the expansion of the CME flux rope, thus containing the larger fraction of mechanical energy; then, as the front plasma is dragged by the expanding flux rope and progressively accelerated, its mechanical energy progressively increases with time, reaching a final stage where both core and front contain approximately the same amount of mechanical energy.

\section{Discussions and Conclusions}
The purpose of our work is to to measure the radial velocity distribution of the CME plasma in the POS. The radial velocity distribution of the CME plasma can provide an important information, while combining the VL and UV channels coronagraph observations. These observations will be provided in the next few years by the Metis coronagraph on-board Solar Orbiter, and by the LST coronagraph on-board the ASO-S. The results presented here will be really useful for the diagnostics of the temperature distribution of CMEs, in combination with the temperature diagnostics discussed by \citet{Bemporad2018}. In this work, we develop a new method based on the cross-correlation method to measure the 2D radial speed distribution of the CME body through synthetic VL coronagraphic images from the MHD simulation. While \citet{Tappin1999} only applied this method to calculate the outflow speeds of the solar wind without considering the variations of speeds at different position angles (and structures) in the images. Comparing with the actual radial velocity from the MHD simulation, we can estimate the measured error, and thus apply this method to a real event. Main results are listed below:

1. We successfully use the synthetic VL images to measure the radial velocity distribution of the CME in the POS. Using two kinds of 2D radial reference velocities of the CME obtained from the MHD simulation (one is integrated along the LOS and averaged with density, the other one is the cut of radial velocity in the POS), we compare the measured radial velocity with these two reference velocities and estimate the measured error between these two velocity distributions. The result shows that the measured velocity obtained from the synthetic VL images is well matched with the reference velocity integrated along the LOS and averaged with the electron number density in the core with no more than $20 \%$ uncertainty, while it is matched better with the simulated radial velocity in the POS in the front within $30 \%$  uncertainty. On the other hand large uncertainties are present in the CME's flanks.

2. We apply the same method to a real CME event observed by STEREO A/COR1 propagating with a low speed of $\sim$320 km s$^{-1}$ at the nose. The 2D radial velocity distribution of this CME shows the anisotropic kinematic feature of the CME plasma when this CME is propagating along the POS. The major velocity of the CME body is slower than the speed of the CME nose, while some parts, especially in the CME flank, have higher speed which is still less than $\sim$400 km s$^{-1}$. 

3. Owing to the low speed of the CME, here, we could estimate the 2D distribution of the Doppler dimming factors for the UV H I Ly-$\alpha$ spectral line, because this factor is really sensitive to the radial component of outflow velocity. If the plasma speed is larger than 300 km s$^{-1}$, this factor is almost zero. This character of the Doppler dimming factor indirectly confirms that \citep[as recently shown by][]{Bemporad2018} the UV intensity in the CME front will be severely attenuated due to the larger plasma radial speed.

4. We derived for the first time the 2D kinetic energy distribution with a three-component structure under the assumption that longitudinal speed is equal to the latitudinal speed, while the latitudinal speed is measured by the slices tracked in the VL images. We also discuss the 2D mechanical energy distribution and time evolution (combining the kinetic and potential energies) in the different parts (core and front) of the CME. For this low speed CME, the potential of the CME is one order of magnitude higher than the kinetic energy. The increase of the kinetic energy is due to both the mass and the speed increase. In the early phases the mechanical energy of the CME core is larger than that of the CME front; then, with time evolution, the front's energy reaches up to the energy of the core. This could be a signature of the CME formation processes, if the CME is originally driven by the expansion of the flux rope leading to subsequent acceleration of plasma embedded in the front.

5. During the CME propagation, the mass of the core almost keeps constant, while that of the CME front increases with the rising of the CME. This is consistent with the classical picture for the formation of the three-part structure of the CME: the dark cavity corresponds to the whole flux rope, and the core, which is more magnetically isolated, is not increasing significantly its mass during the CME expansion, while the front forms by the mass supplement from the dimming region in the low corona and/or pile-up of the ambient plasma, leading to a progressive mass increase.

In our work, we have compared the difference between the kinetic energy with and without considering the latitudinal and longitudinal speed of the CME. The comparison ratio between these two kinds of kinetic energies is no more than 10$\%$ during the evolution of the CME from 11:25 to 12:50 UT. Given the mechanical energy evolution and distribution of the CME core and front, the influence made by these two additional kinetic energy components is thus negligible, because the energy difference between these two kinetic energies is no more than 2$\%$ of the total mechanical energy. The largest uncertainty of the energy is related with the estimate of the CME mass, that was performed here with two different background subtractions, providing upper and lower limit estimates of the CME mass. This is important because MHD simulations \citep{Lugaz2005} and other works dealing with the CME mass measurement \citep[e.g.][]{Vourlidas2000,Vourlidas2010, Howard2015b} have shown that the estimate of the CME masses based on Thomson scattering method may be underestimated by a factor of $\sim 2$.

Doppler dimming factor $F_D$ plays a significant role in calculating the radiative emission due to the resonant scattering of the chromospheric H I Ly-$\alpha$ emission by very few remaining neutral hydrogen atoms in the hot corona. This parameter will be affected by the radial velocity of the plasma, the normalized coronal absorption profile, and the intensity spectrum of incident profile, as mentioned above. In our work, we do not consider the velocity projection effect due to the propagation of the real CME along the POS. It is possible to correct the projection effect via the polarization-ratio technique mentioned by \citet{Susino2016}. The purpose of our work is to provide a good tool to derive the 2D radial velocity distribution of the CME, so that in the future we could combine the VL and UV coronagraphic observations, thanks to the future Solar Orbiter/Metis and ASO-S/LST instruments, to diagnose the characteristics of the CMEs, such as the kinetic energy distribution, temperature distribution, and so on. \citet{Bemporad2018} tested three different hypothesis to estimate the CME plasma temperatures by taking the Doppler dimming factor into account: (1) full collisional excitation assumption, (2) full radiative excitation assumption and (3) radiative and collisional excitation assumption. The results in \citet{Bemporad2018} showed that all temperature estimated by the synthetic VL and UV images were underestimated, whatever method is considered to take into account LOS integration effects.

As we all know, the Sun is variable, constantly evolving in time. For the UV intensity observations from the future coronagraphs, whether there are active region or not in the solar disk will lead to an immeasurable different intensity measurement of the resonant radiative contributions from the chromosphere. In the era of ASO-S, LST will first supply the full disk observation of  the H I Ly-$\alpha$ line, which can allow us to constrain the incident radiation intensity from the chromosphere layer, and enhance the accuracy of the calculation of the Doppler dimming factor, subsequently decreasing the uncertainty of the CME temperature estimation.

A good approach to understand the dynamical evolution of the physical parameters of the erupting CME is to analyse the CME energy budget. The energy of the CME propagating outward consists of the kinetic, potential, magnetic and other energy forms. Many authors have found that there should exist additional heating sources to heat CMEs, to reproduce the plasma temperatures as derived from UV emission and taking into account the CME plasma cooling mainly due to the adiabatic expansion. The corresponding thermal energy can be comparable or even higher than kinetic and potential energies carried by the CMEs \citep{Akmal2001, Murphy2011, Lee2009, Landi2010}. Furthermore, the temperature increase of the plasma during the early CME expansion phase also implied the existence of an additional thermal-energy source \citep{Bemporad2007}. If CMEs are assumed to be a perfect gas in local thermodynamic equilibrium with equal electron and ion temperatures, the enthalpy can be as large as five times thermal energy. The upper limit of the enthalpy of a normal CME will be one order of magnitude less than the kinetic and potential energies, except in the lower corona where it can be comparable to the kinetic energy \citep{Vourlidas2000}. For fast CMEs the magnetic energy is negligible compared to the potential and kinetic energy, while slow CMEs could be driven by magnetic force. Hence, estimate of the 2D temperature distribution inside CMEs will reveal the real evolution of the CME thermal energy during their expansion. The total energies of the CMEs are comparable with the range of flare energies estimated from non-thermal electrons \citep{Aschwanden2002, Emslie2004, Emslie2005, Feng2013}. In other word, eruptive events possibly share a relationship of an energy-equal partition between CMEs and flares. Furthermore, combining with the analysis of the energy partition between the flares and CMEs among the various components would provide valuable constraints on the fundamental energy release process(es) \citep{Reeves2010a, Reeves2010b}, thus helping to understand the origin of CMEs.




\acknowledgements
Thanks to the host of Dr Thomas Wiegelmann and Dr Bernd Inhester at the Max-Planck-Institut for Solar System Research, laying the foundation for the preparation of the article. STEREO is a project of NASA. The SECCHI data used here were produced by an international consortium of the Naval Research Laboratory (USA), Lockheed Martin Solar and Astrophysics Lab (USA), NASA Goddard Space Flight Center (USA), Rutherford Appleton Laboratory (UK), University of Birmingham (UK), Max-Planck-Institut for Solar System Research (Germany), Centre Spatiale de Li{\`e}ge (Belgium), Institut d'Optique Th{\'e}orique et Applique{\'e} (France), Institut d'Astrophysique Spatiale (France).  This work is supported by NSFC (Grant Nos. 11522328, 11473070, 11427803 and U1731241), by CAS Strategic Pioneer Program on Space Science (Grant Nos. XDA15010600, XDA15052200, XDA15320103 and XDA15320301), and by the National Key R\&D Program of China (2018YFA0404200). This research has received funding from the the European Union Horizon 2020 research and innovation programme (grant agreement No. 647214). This work used the DiRAC@Durham facility managed by the Institute for Computational Cosmology on behalf of the STFC DiRAC HPC Facility (www.dirac.ac.uk). The equipment was funded by BEIS capital funding via STFC capital grants ST/P002293/1, ST/R002371/1 and ST/S002502/1, Durham University and STFC operations grant ST/R000832/1. DiRAC is part of the National e-Infrastructure.
We acknowledge the use of the open source (gitorious.org/amrvac) MPI-AMRVAC software, relying on coding efforts from C. Xia, O. Porth, R. Keppens.

\bibliography{refs}
\end{document}